%% file: main.tex
\documentclass[11pt, letterpaper]{article}

\usepackage[left=1in, right=1in, top=1in, bottom=1in]{geometry}
\usepackage[utf8]{inputenc}
\usepackage[T1]{fontenc}
\usepackage{bm}
\usepackage{type1cm}
\usepackage{lettrine}
\usepackage{amsmath,amssymb,amsthm}
\usepackage{moreverb}
\usepackage{mathtools}
\usepackage{amsmath}
\usepackage{amssymb}
\usepackage{algorithmic}
\usepackage{graphics}
\usepackage{graphicx}
\usepackage{caption}
\usepackage{extarrows}
\usepackage{color}
\usepackage{framed}
\usepackage{wrapfig}
\usepackage{bm}
\usepackage{mathrsfs}
\usepackage{mathabx}
\usepackage{multirow}
\usepackage{longtable}
\usepackage{hyperref}
\usepackage{paralist}
\usepackage{indentfirst}
\usepackage{relsize}
\usepackage{extarrows}
\usepackage{upgreek}
\usepackage{bm}
\usepackage{mwe}
\usepackage[dvipsnames,table]{xcolor}
\usepackage{booktabs}
\usepackage{authblk}
\usepackage{lettrine}
\usepackage{type1cm}
\usepackage{threeparttable}
\usepackage[sort&compress,numbers]{natbib}
\usepackage[figurename=Figure]{caption}
\usepackage[normalem]{ulem}
 
\graphicspath{ {./figures/} }
\usepackage[font=footnotesize,labelfont=bf]{caption}
\providecommand{\keywords}[1]{\textbf{\textit{Keywords: }} #1}

\usepackage{authblk}
\usepackage{rotating}
\usepackage{subcaption}
\usepackage{afterpage}

\hypersetup{
bookmarksopen=true,
bookmarksnumbered=true,
unicode=false,
pdftoolbar=true,
pdfmenubar=true,
pdffitwindow=false,
pdfstartview={FitH},
pdftitle={My title},
pdfauthor={Author},
pdfsubject={Subject},
pdfcreator={Creator},
pdfproducer={Producer},
pdfkeywords={keywords},
pdfnewwindow=true,
colorlinks=true,
linkcolor=blue,
citecolor=blue,
filecolor=blue,
urlcolor=blue
}

\begin{document}

\title{\textbf{Tractogram foundation model}}

\author[1,2,3,*]{Guikun Chen}  
\author[4,*]{Yuqian Chen}  
\author[5]{Yijie Li}  
\author[6,3]{Yogesh Rathi}  
\author[7,3]{Nikos Makris}  
\author[5,\dag]{\\Fan Zhang}  
\author[1,\dag]{Wenguan Wang}  
\author[2,3,\dag]{Lauren J. O'Donnell}  

\affil[1]{\small The State Key Lab of Brain-Machine Intelligence, Zhejiang University, Hangzhou 310058, Zhejiang, China}
\affil[2]{\small Department of Radiology, Brigham and Women's Hospital, Mass General Brigham, Boston 02115, Massachusetts, USA}
\affil[3]{\small Harvard Medical School, Boston 02115, Massachusetts, USA}
\affil[4]{\small Academy of Medical Engineering and Translational Medicine, Tianjin University, Tianjin 300072, China}
\affil[5]{\small School of Information and Communication Engineering, University of Electronic Science and Technology of China, Chengdu 610054, Sichuan, China}
\affil[6]{\small Psychiatry Neuroimaging Laboratory, Brigham and Women's Hospital, Mass General Brigham, Boston 02115, Massachusetts, USA}
\affil[7]{\small Department of Psychiatry, Center for Morphometric Analysis, Massachusetts General Hospital, Boston 02115, Massachusetts, USA}

\vspace{8pt}

\affil[ ]{\textsuperscript{*}These authors contributed equally \quad \textsuperscript{\dag}Corresponding authors}

\date{}

\maketitle

\normalsize

\input{sec/abs.tex}


\input{sec/intro.tex}

\input{sec/results.tex}
\input{sec/discussion.tex}
\input{sec/methods.tex}

\section*{Data availability}
The source diffusion MRI datasets used in the study, i.e., HCP~\citep{van2013wu}, ABIDE-II~\citep{di2014autism}, ADNI~\citep{mueller2005alzheimer}, CNP~\citep{poldrack2016phenome}, and PPMI~\citep{marek2011parkinson}, are publicly available from the original data repositories upon approved application, together with cohort-specific acquisition protocols and imaging metadata. Tractograms were generated from the released diffusion MRI data using three tractography pipelines: UKF tractography, available through SlicerDMRI (\href{https://dmri.slicer.org}{https://dmri.slicer.org}) and the UKF tractography repository (\href{https://github.com/pnlbwh/ukftractography}{https://github.com/pnlbwh/ukftractography}); DSI-Studio tractography (\href{https://dsi-studio.labsolver.org}{https://dsi-studio.labsolver.org}); and iFOD2 tractography implemented in MRtrix3 (\href{https://www.mrtrix.org}{https://www.mrtrix.org}).

\bibliographystyle{unsrt}
\bibliography{sn-bibliography}

\end{document}

%% file: sec/abs.tex
\begin{abstract}
\small

Diffusion MRI (dMRI) tractography is the only noninvasive approach for mapping white-matter pathways in the living human brain. It represents each brain as a tractogram: a large, unordered set of three-dimensional streamlines that includes information about both local streamline geometry and whole-brain anatomical organization. This structure makes tractograms a natural but challenging target for representation learning. Existing methods treat streamline classification and subject-level prediction as separate problems: streamline classifiers focus on geometric patterns, whereas subject-level prediction often depends on hand-crafted features. As a result, current methods do not learn reusable representations that connect streamline anatomy with whole-brain inter-subject variation. Here we introduce TractFM, a tractogram foundation model that learns reusable representations directly from whole-brain streamline sets. TractFM combines a local streamline encoder with a permutation-equivariant tractogram encoder, allowing all streamlines from a subject to be contextualized jointly in a single forward pass. Pretraining on dense anatomical tract parcellation, i.e., assigning anatomical labels to individual streamlines, yields two complementary representations: contextualized streamline-level embeddings for tract parcellation and compact subject-level descriptors for downstream prediction of subject phenotypes. Across three tractography algorithms and five dMRI datasets, TractFM transfers to both streamline-level and subject-level tasks. Its frozen representations achieve accurate tract parcellation and predict age and sex across independent datasets. These results show that whole-brain geometric context, learned once, can generalize across tractography pipelines, datasets, and prediction tasks. More broadly, TractFM demonstrates the feasibility of reusable, transferable representations from geometric neuroimaging data, extending the foundation model paradigm beyond intensity-based medical images.

\end{abstract}

%% file: sec/intro.tex
\section*{Introduction}
\label{sec:intro}

Diffusion MRI tractography is the only non-invasive imaging technique that can delineate white-matter pathways in vivo~\cite{jeurissen2019diffusion,zhang2022quantitative}. These pathways are represented in a tractogram, which is an unordered set of thousands to millions of variable-length three-dimensional curves (streamlines). Tractograms provide a subject-specific geometric reconstruction of structural brain connections that can support large-scale connectomic studies relating brain structure to behavior, disease, and treatment response~\cite{hagmann2008mapping,smith2012anatomically,thomas2014anatomical,campbell2014potential,calabrese2015diffusion,whitfield2016brain}. Analysis of these reconstructions, however, remains difficult: tractograms are high-dimensional, contain complex anatomical information at multiple scales, including the shape of individual streamlines and the global configuration of pathways within the brain, and are sensitive to acquisition and processing choices~\cite{maier2017challenge,sarwar2019mapping,schilling2019limits,yeh2021mapping,zhang2025think}.

These challenges make tractography a natural setting for foundation-model-style representation learning. In image-based medical AI, large-scale pretraining has produced representations that transfer across tasks, datasets, and clinical settings~\cite{moor2023foundation,paschali2025foundation,pai2024foundation,wu2025towards,shao2025mri,zhang2025foundation}. A similar paradigm would be valuable for tractography: a model could learn reusable representations of whole-brain white-matter organization once, then transfer them to both streamline-level anatomical labeling and subject-level prediction. Yet this paradigm has not reached tractography, largely because tractograms are unordered geometric point sets rather than the intensity grids assumed by most medical imaging architectures. Tractography, therefore, requires models that are geometry-aware, permutation-equivariant, and capable of reasoning about the tractogram as a whole~\cite{moor2023foundation,paschali2025foundation,zhang2025think}. Recent reviews have explicitly called for adaptive, generalizable modeling frameworks for tractography that can operate across protocols, populations, and downstream tasks~\cite{descoteaux2025millennium,zhang2025think}.

Tract parcellation, which assigns streamlines to anatomically meaningful bundles, is the central anchoring task in supervised learning for tractography and a prerequisite for most tract-specific quantification~\cite{zhang2018anatomically,zhang2020deep,xue2023tractcloud}. Most existing parcellation methods classify streamlines individually using local geometric descriptors, effectively treating the tractogram as a bag of independent samples~\cite{xu2019objective,wang2019dynamic,zhang2020deep}. More recent methods add broader anatomical context by sampling a limited number of local and global streamlines~\cite{xue2023tractcloud,bisten2026rapidparc}.
However, tractography analysis also requires subject-level representations that summarize inter-subject variation in whole-brain white-matter organization. Such representations are essential for population-level studies and phenotypic prediction, where variables such as age and sex provide biologically meaningful tests of whether a model captures subject-level anatomical structure. Existing subject-level tractography models typically rely on manually designed tractography-derived features~\cite{mwangi2013prediction,he2022model,chen2023tractgraphcnn,chen2025tractgraphformer}. As a result, tractography representation learning remains fragmented: existing methods either focus on dense streamline labeling or on task-specific subject-level prediction, but do not learn reusable representations that connect streamline anatomy with whole-brain inter-subject variation.

We present TractFM, a tractogram foundation model that operates directly on the whole-brain streamline set. TractFM ingests an entire subject's tractogram in a single forward pass and resolves labels for all streamlines jointly, so that each streamline is evaluated within the full structural configuration while retaining its individual geometry. We adopt dense tract parcellation as the pretraining objective because it is anatomically grounded and inherently context-dependent, so a model that succeeds at whole-brain parcellation must encode contextualized features reflecting both streamline geometry and tractogram-wide organization~\cite{reisert2011global,smith2012anatomically}. These are precisely the features that subject-level tractography analyses require. The same forward pass therefore yields two outputs from one pretrained model: dense anatomical parcellation and a compact subject-level embedding that can be reused across downstream phenotypic prediction tasks without encoder adaptation.

We validate TractFM across multiple independent cohorts to test whether this representation is both anatomically precise and biologically informative. For dense tractography analysis, TractFM achieves accurate whole-brain tract parcellation across datasets, demonstrating that whole-brain contextual modeling can support streamline-level anatomical labeling. For subject-level transfer, we freeze the TractFM encoder and show that its learned embeddings retain phenotypic information sufficient to predict chronological age and sex without adapting the encoder to these downstream tasks. Such a capability directly addresses the field's need to capture and account for structural white matter variability across populations and the lifespan~\cite{descoteaux2025millennium}. Together, these experiments establish TractFM as a unified representation-learning framework for tractography, linking dense anatomical parcellation with reusable subject-level embeddings. By modeling the full geometric structure of the tractogram, TractFM can provide a foundation for tractography analysis across clinical neuroimaging and population neuroscience.

%% file: sec/results.tex
\input{sec/misc/fig_method}

\section*{Results}
\label{sec:res} 

\subsection*{TractFM models tractogram at the whole-brain level}
Current deep learning methods for diffusion MRI tractography typically represent streamlines either independently~\cite{xu2019objective,wang2019dynamic,zhang2020deep} or with limited contextual information, such as local neighborhoods or a reduced set of globally sampled references~\cite{xue2023tractcloud,bisten2026rapidparc}. These designs can perform the task for which they are trained, but they do not provide a general tractogram representation: single-streamline models lack access to whole-brain context, context-sampling models observe only a subset of the tractogram, and subject-level predictors are usually optimized directly for manually designed features rather than exposing reusable anatomical embeddings~\cite{mwangi2013prediction,he2022model,chen2023tractgraphcnn,chen2025tractgraphformer}. This work develops TractFM as a tractogram foundation model in this operational sense: it is pre-trained on an anatomically grounded, dense whole-brain objective and, after removal of the task-specific head, provides frozen streamline- and subject-level embeddings that can be reused for downstream analyses without encoder adaptation. The model follows a two-stage design (cf. Fig.~\ref{fig:method_overview}): a local streamline encoder first extracts orientation-invariant geometric features for each streamline, and a tractogram encoder then uses dense self-attention to refine these features in the context of the full streamline set.

This whole-brain formulation serves two distinct but synergistic objectives. First, it enables dense anatomical parcellation in a single forward pass, dynamically mapping individual streamlines to both anatomical fiber tract labels and fine-grained fiber clusters within anatomical tracts. Second, after removing the task-specific classification head, the pre-trained model yields one whole-brain-contextualized embedding per streamline. Simple summary statistics of these embeddings can then be combined into a compact, fixed-dimensional subject-level fingerprint. This separation between pre-training, representation extraction, and downstream readout is what distinguishes TractFM from previous task-specific tractography models and motivates evaluating it as a foundation model. As demonstrated in subsequent sections, this frozen, subject-level representation is transferred to downstream phenotypic prediction across multiple independent datasets.

\subsection*{TractFM jointly encodes streamlines within the whole-brain tractogram}

\input{sec/misc/tab_parcel}

To validate the representational capacity of TractFM, we first evaluated its performance on the pre-training task: dense anatomical parcellation. The model was benchmarked on a task of classifying streamlines into 43 anatomical fiber tracts. To rigorously assess geometric robustness, all models were evaluated both on the original subject space and on a synthetically transformed (STA) dataset incorporating random global spatial perturbations (rotations, translations, and scaling)~\cite{xue2023tractcloud}.

Table~\ref{tab:parcellation} summarizes the parcellation performance against established baselines. Traditional single-streamline architectures achieve competitive accuracy ($\sim$90-91\%) on the original data. However, because these networks process each streamline without reference to the rest of the tractogram, they exhibit severe vulnerability to spatial perturbations, with their accuracy degrading by up to 10\% on the STA dataset. While integrating randomly sampled context~\cite{xue2023tractcloud} mitigates this degradation, TractFM establishes a new state-of-the-art performance across both configurations. By leveraging whole-brain global attention, TractFM achieved an accuracy of 93.11\% and a macro-F1 score of 91.11\% on the original data. In addition, TractFM demonstrated spatial resilience, maintaining 92.48\% accuracy on the STA dataset, thereby demonstrating that comprehensive relative contextualization effectively eliminates the need for spatial registration.

To examine the contribution of global context more directly, we removed the Transformer-based tractogram encoder and retained only the local streamline encoder (TractFM w/o context). This ablation substantially reduced performance. On the original data, accuracy decreased from 93.11\% to 84.77\%, and macro-F1 decreased from 91.11\% to 80.13\%. The same pattern was observed on the STA dataset, where macro-F1 fell from 90.19\% to 76.30\%. The larger drop in macro-F1 than in accuracy suggests that global context is particularly important for anatomically difficult or underrepresented classes. In other words, local streamline geometry alone may be sufficient for some canonical pathways, but accurate disambiguation of spatially similar or crossing bundles benefits from reference to the configuration of the full tractogram.

\subsection*{TractFM embeddings recapitulate anatomical organization}

\input{sec/misc/fig_tsne}

We next asked whether TractFM learns streamline embeddings that reflect tract-level anatomical organization rather than only supporting the supervised parcellation task. Using frozen contextualized streamline embeddings from an unseen test subject (cf. Fig.~\ref{fig:tsne_embeddings}), we observed clear tract-specific structure across streamlines from 42 representative tract classes. In the native 128-dimensional space, the embeddings achieved a silhouette score of $0.349$, compared with a permutation-based null distribution of $-0.107 \pm 0.007$ (permutation $p = 9.99 \times 10^{-4}$). Local neighborhood structure was also preserved, with $k$-nearest-neighbor purity of $94.8\%$, $91.5\%$, and $86.2\%$ at $k = 5$, $10$, and $15$, respectively (all permutation $p = 5.00 \times 10^{-4}$).

This organization remained evident after visualization in two dimensions. The t-SNE projection showed a modular arrangement of white-matter pathways, with a 2D silhouette score of $0.592$ compared with a permutation-based null distribution of $-0.255 \pm 0.026$. In the 2D embedding, $k$-nearest-neighbor purity reached $95.5\%$, $93.0\%$, and $88.9\%$ at $k = 5$, $10$, and $15$, respectively, and the projection retained high trustworthiness with respect to the native embedding space ($0.998$ at $k = 15$). Together, these analyses indicate that the learned representation preserves tract-level anatomical organization rather than merely encoding low-level geometric similarity.

To examine how the model uses whole-brain context, we extracted dense multi-head attention matrices from the final tractogram encoder block and visualized the attention profile of a representative query streamline from the corpus callosum 5 (CC5) in a separate, unseen test subject (cf. Fig.~\ref{fig:attn_map}). The resulting attention map was sparse and strongly non-uniform, concentrating on a restricted subset of streamlines rather than being diffusely distributed across the tractogram. In this example, the highest-weighted context streamlines arose from SLF-I, CC4, superficial frontal--parietal streamlines, and CC5, suggesting that the model integrates both within-tract and cross-tract contextual information when encoding a target streamline.

Overall, these results support the view that TractFM learns a structured latent space aligned with white-matter anatomy and uses attention selectively rather than uniformly. This provides a plausible basis for transferring frozen representations to downstream cohorts.

\afterpage{\clearpage}
\input{sec/misc/fig_attn}
\clearpage

\subsection*{Zero-tuning transfer supports downstream phenotype prediction}
\label{sec:zero_tuning_multisite}

We next evaluated whether TractFM learns a reusable structural representation that transfers beyond the anatomical parcellation pre-training task. After pre-training on atlas tractograms derived from the HCP Unrelated 100 cohort (a curated subset of the HCP Young Adult (HCP-YA) release consisting of 100 subjects with no family relations), we froze the foundation model and trained only lightweight downstream predictors on subject-level descriptors derived from the contextualized streamline embeddings. Specifically, for each subject we summarized the matrix of 128-dimensional frozen streamline embeddings with permutation-invariant statistics across streamlines, including the mean, standard deviation, covariance eigenvalues, and spectral entropy, to obtain fixed-dimensional descriptors. Throughout the downstream analyses and tables, descriptor labels use compact capitalized abbreviations: Mean denotes the per-dimension embedding mean, Std the per-dimension standard deviation, CovEig the covariance eigenvalues, and Entropy the spectral entropy; compound labels indicate concatenation of the corresponding summaries, as in Mean+Std+CovEig+Entropy. Transfer was evaluated in two increasingly challenging settings: the broader HCP-YA S1200 release, which includes the Unrelated 100 subset, and the ABIDE-II+ADNI+CNP+PPMI cohort, which spans distinct sites, acquisition protocols, age ranges, and clinical populations. Because the TractFM encoder was not updated during downstream model fitting, these analyses evaluate zero-tuning use of the pre-trained representation. The S1200 analysis tests task transfer within the same broader HCP-YA release, whereas the ABIDE-II+ADNI+CNP+PPMI benchmark additionally introduces substantial variation in site, acquisition, demographics, and disease composition.

In the lower-shift HCP-YA S1200 setting, frozen TractFM embeddings supported accurate phenotype prediction without encoder adaptation (Table~\ref{tab:zerotuning_overview}). For age, the best-performing configuration used the Mean+Std descriptor with Ridge regression, achieving an MAE of 2.799 years and Pearson $r = 0.405$. Given the restricted age range in HCP-YA, this result is better interpreted as successful rank-order prediction under a compressed dynamic range than as lifespan-scale age regression. For sex, the Mean descriptor combined with logistic regression yielded the strongest overall performance, with accuracy $= 0.889$, AUC $= 0.954$, and F1 $= 0.889$. These findings indicate that even simple summary statistics of frozen TractFM embeddings retain substantial phenotype-relevant information in a relatively homogeneous transfer setting.

\input{sec/misc/tab_downstream1}

Transfer became more challenging in the ABIDE-II+ADNI+CNP+PPMI cohort, where the encoder remained frozen, and the same predefined subject-level descriptors were applied without adaptation to the target domain. Nevertheless, the embeddings remained predictive for both age and sex (Table~\ref{tab:multisite_feature_hierarchy}). For age, the Mean descriptor with SVR achieved MAE $= 5.878$ years and Pearson $r = 0.958$. Adding embedding-dispersion information improved performance further: incorporating the per-dimension standard deviation reduced the error to MAE $= 5.440$ years and increased Pearson correlation to $r = 0.962$. The 385-dimensional Mean+Std+CovEig+Entropy descriptor, which combines the embedding mean, standard deviation, covariance eigenvalues, and spectral entropy, produced the best overall age prediction (MAE $= 5.360$ years, $r = 0.962$). By contrast, covariance eigenvalues alone were weaker (MAE $= 7.755$ years, $r = 0.909$), suggesting that higher-order covariance structure is useful when combined with first-order embedding information rather than used in isolation.

A similar pattern was observed for sex classification in the ABIDE-II+ADNI+CNP+PPMI cohort. The Mean descriptor generalized across datasets, reaching accuracy $= 0.720$, AUC $= 0.782$, and F1 $= 0.779$. Adding embedding-dispersion statistics improved all major metrics (Mean+Std descriptor: accuracy $= 0.760$, AUC $= 0.826$, F1 $= 0.812$). Incorporating spectral structure yielded a further gain, with the Mean+Std+CovEig+Entropy descriptor achieving the highest accuracy, AUC, and F1 (accuracy $= 0.763$, AUC $= 0.835$, F1 $= 0.813$). Again, covariance eigenvalues alone performed worse than descriptors that preserved first-order embedding information (accuracy $= 0.678$, AUC $= 0.753$, F1 $= 0.720$). Together, these results indicate that higher-order covariance summaries capture additional phenotypically relevant variation beyond the embedding mean alone, with the benefit becoming most apparent under stronger domain shift.

Across downstream models, the frozen embedding space was sufficiently organized that simple linear downstream models (e.g., Ridge regression) remained competitive (Table~\ref{tab:multisite_model_comparison}). On the ABIDE-II+ADNI+CNP+PPMI cohort, Ridge regression applied to the Mean+Std+CovEig+Entropy descriptor achieved MAE $= 6.455$ years and $r = 0.952$ for age, while logistic regression reached accuracy $= 0.726$, AUC $= 0.817$, and F1 $= 0.773$ for sex. Nonlinear models improved performance further, with SVR and RBF-SVC yielding the strongest overall results. However, the gap between linear and nonlinear models was modest relative to the overall transfer performance, suggesting that a major fraction of phenotype-related signal is already accessible in the frozen TractFM representation.

For qualitative context, prior structural-connectome studies have reported age-prediction performance in related settings, including correlations around $r = 0.79$ with roughly 9 years average deviation in 201 healthy subjects, age prediction at $r = 0.21$ with MAE $= 3.02$ years in 1,048 HCP subjects, and elderly-cohort brain-age correlations of $r = 0.535$, $0.529$, and $0.517$ for streamline counts, SIFT2, and ReAl-LiFE, respectively \citep{han2014predicting,kopetzky2024predictability,gurusamy2020diffusion}. Because those studies differ in cohort composition, age range, tractography pipeline, preprocessing, and evaluation protocol, they are included here only as broad context rather than as direct benchmarks.

\subsection*{Whole-brain inference remains computationally practical}

A potential concern for whole-brain modeling is whether jointly encoding 10K streamlines with dense self-attention is computationally practical. In end-to-end evaluation on the test split of the HCP Unrelated 100 pre-training dataset, comprising 20 subjects with 30 tractography augmentations (random rotations, translations, and scaling) per subject (620 tractograms in total), TractFM completed whole-brain parcellation, that is, whole-tractogram encoding via self-attention blocks followed by streamline classification via an MLP-based prediction head, in 5 min 8 s, whereas TractCloud required 19 min 25 s. This corresponds to a $\sim3\times$ speedup in total wall-clock time. Averaged over all subject-augmentation pairs, the runtime decreased from 1.88 s to 0.50 s per tractogram.

To isolate raw inference behavior, we additionally benchmarked both models on random inputs under their default practical settings. TractCloud was evaluated with its standard batch size of 1,024 streamlines, producing output logits in 1.61 s, with peak allocated and reserved GPU memory of 4183.44 MB and 4386.00 MB, respectively. By contrast, TractFM processed a full 10,000-streamline tractogram in a single forward pass, producing output logits in 0.25 s, with peak allocated and reserved GPU memory of 6242.70 MB and 6522.00 MB, respectively. Thus, despite operating on nearly an order of magnitude more streamlines per inference, TractFM was $\sim 6\times$ faster in wall-clock time. Importantly, TractCloud could not be practically scaled to comparably large batch sizes in our setting, whereas TractFM remained tractable for whole-brain inference in a single pass.

These results define the practical operating point examined here. As expected from dense global attention, TractFM incurred a higher peak memory footprint, but this cost remained manageable at 10K streamlines and enabled substantially faster inference by avoiding repeated small-batch processing. More broadly, the proposed formulation is not intrinsically tied to a fixed number of streamlines. Although processing an approximated set of 10K streamlines is currently a pragmatic choice dictated by available hardware, TractFM treats the tractogram as a permutation-equivariant set and is therefore conceptually agnostic to $N$ (number of streamlines that represents the whole brain). This makes the framework naturally positioned to benefit from future advances in memory capacity and more efficient attention mechanisms, potentially extending whole-brain contextual modeling to substantially denser tractograms without changing the underlying learning paradigm.

\subsection*{Robustness to tractography-method variance and stochastic subsampling}

\input{sec/misc/tab_robust}

We next assessed whether the zero-tuning downstream results were sensitive to two sources of tractogram variability: stochastic subsampling of dense tractograms and the tractography method used to generate streamlines. Because the 385-dimensional Mean+Std+CovEig+Entropy descriptor was the best-performing representation for both age and sex in  ABIDE-II+ADNI+CNP+PPMI analysis, these robustness analyses used the same cohort split, descriptor, and downstream models selected in the primary UKF analysis, namely SVR for age prediction and RBF-SVC for sex classification (Table~\ref{tab:multisite_model_comparison}). For stochastic subsampling, each UKF tractogram was represented by three independent 10K-streamline samples, reflecting the practical setting in which a whole-brain tractogram typically contains more than 1M streamlines but the model is evaluated on a fixed-size subset. Across the three samples, age prediction varied minimally, with MAE ranging from 5.356 to 5.434 years and Pearson $r$ from 0.962 to 0.963 (Table~\ref{tab:robustness_sampling}). Sex classification also showed limited variation, with accuracy from 0.763 to 0.775, AUC from 0.835 to 0.849, and F1 from 0.813 to 0.825. These results indicate limited sensitivity to the specific 10K-streamline sample under the evaluated UKF protocol.

We then evaluated robustness across tractography methods by applying the frozen-embedding downstream pipeline to UKF, DSI Studio, and iFOD2 tractograms. Age prediction remained consistent across methods, with MAE ranging from 5.050 to 5.715 years and Pearson $r$ from 0.958 to 0.967 (Table~\ref{tab:robustness_tractography}). Sex classification showed limited variation in accuracy, AUC, and F1, with accuracy ranging from 0.737 to 0.764, AUC from 0.820 to 0.844, and F1 from 0.778 to 0.817. These results indicate that the frozen TractFM representation was stable across the evaluated tractography methods.

%% file: sec/misc/fig_method.tex
\begin{figure}[t]
    \centering
    \includegraphics[width=\linewidth]{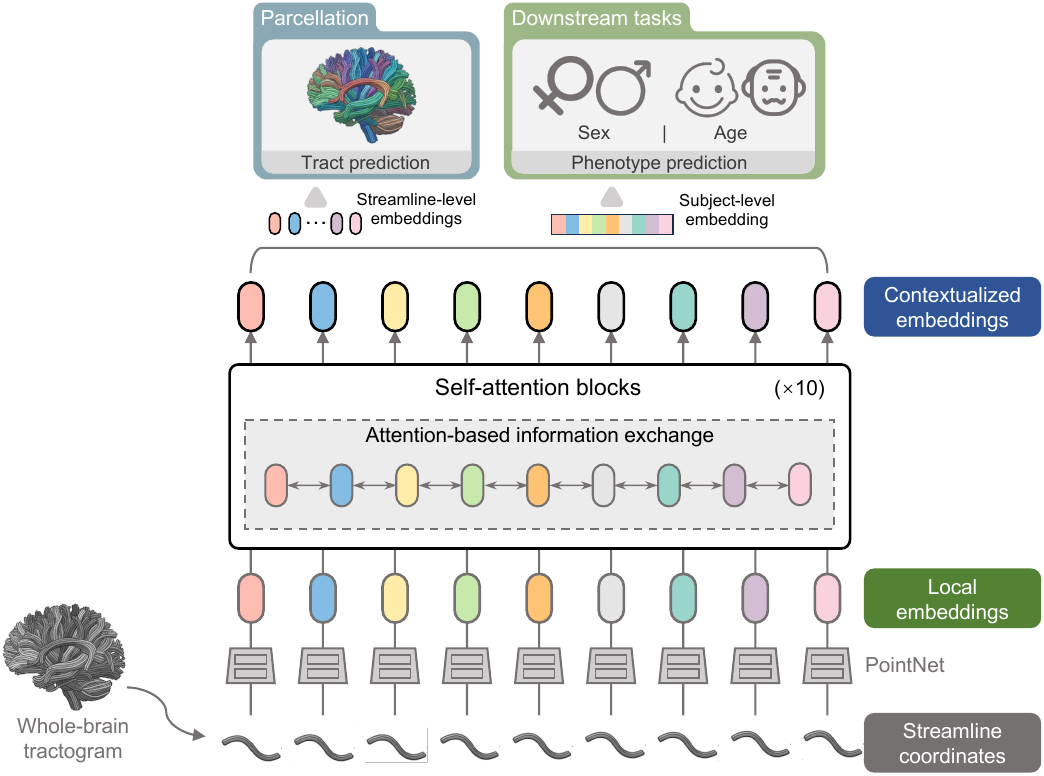}
    \caption{\textbf{Overview of the TractFM whole-brain tractogram modeling framework.}
    A subject's tractogram is represented as an unordered set of streamline coordinates. A PointNet-style streamline encoder first maps each streamline to a local geometric embedding, and a stack of self-attention blocks then exchanges information across the full streamline set to produce whole-brain-contextualized embeddings. During pre-training, these embeddings are used for dense tract parcellation, assigning anatomical labels to streamlines in a single forward pass. After pre-training, the task-specific head is removed and the frozen contextualized embeddings are summarized into subject-level descriptors for downstream phenotype prediction, including sex classification and age regression.}
    \label{fig:method_overview}
\end{figure}

%% file: sec/misc/tab_parcel.tex
\begin{table*}[!t]
\centering
\caption{\textbf{Anatomical parcellation performance across spatial transformations.} Results are reported on the held-out test cohort for both original coordinate data and synthetically transformed (STA) augmented data. The ablation of TractFM (without context) relies exclusively on the local streamline encoder. Abbreviations: Acc, Top-1 Accuracy; F1, Macro-averaged F1 score.}
\label{tab:parcellation}

\small
\setlength{\tabcolsep}{8pt}

\begin{tabular}{lcccc}
\toprule
\multirow{2}{*}{\textbf{Method}}
& \multicolumn{2}{c}{Original} 
& \multicolumn{2}{c}{STA} \\
\cmidrule(lr){2-3} \cmidrule(lr){4-5}

& Acc (\%) & F1 (\%) 
& Acc (\%) & F1 (\%) \\
\midrule

DeepWMA~\cite{zhang2020deep}    & 90.29 & 88.12  & 82.35 & 76.55 \\
DCNN++~\cite{xu2019objective}    & 91.26 &  89.14  & 84.14  &  79.16 \\

PointNet~\cite{qi2017pointnet}    & 91.36 &  89.12  & 81.83  &  75.89 \\
TractCloud$_\text{PointNet}$\cite{xue2023tractcloud}    & 92.28 &  90.36  & 91.57  & 89.40  \\

DGCNN~\cite{wang2019dynamic}    & 91.85 &  89.78  & 83.70  & 78.55  \\
TractCloud$_\text{DGCNN}$~\cite{xue2023tractcloud}    & 91.99 &  90.10  & 91.69  & 89.65  \\

\midrule

\textbf{TractFM (w/o context)}  & 84.77 & 80.13  & 82.28 & 76.30 \\
\textbf{TractFM (Ours)}  & \textbf{93.11} & \textbf{91.11}  & \textbf{92.48} & \textbf{90.19} \\

\bottomrule
\end{tabular}

\end{table*}

%% file: sec/misc/fig_tsne.tex
\begin{figure}[t]
    \centering
    \includegraphics[width=\linewidth]{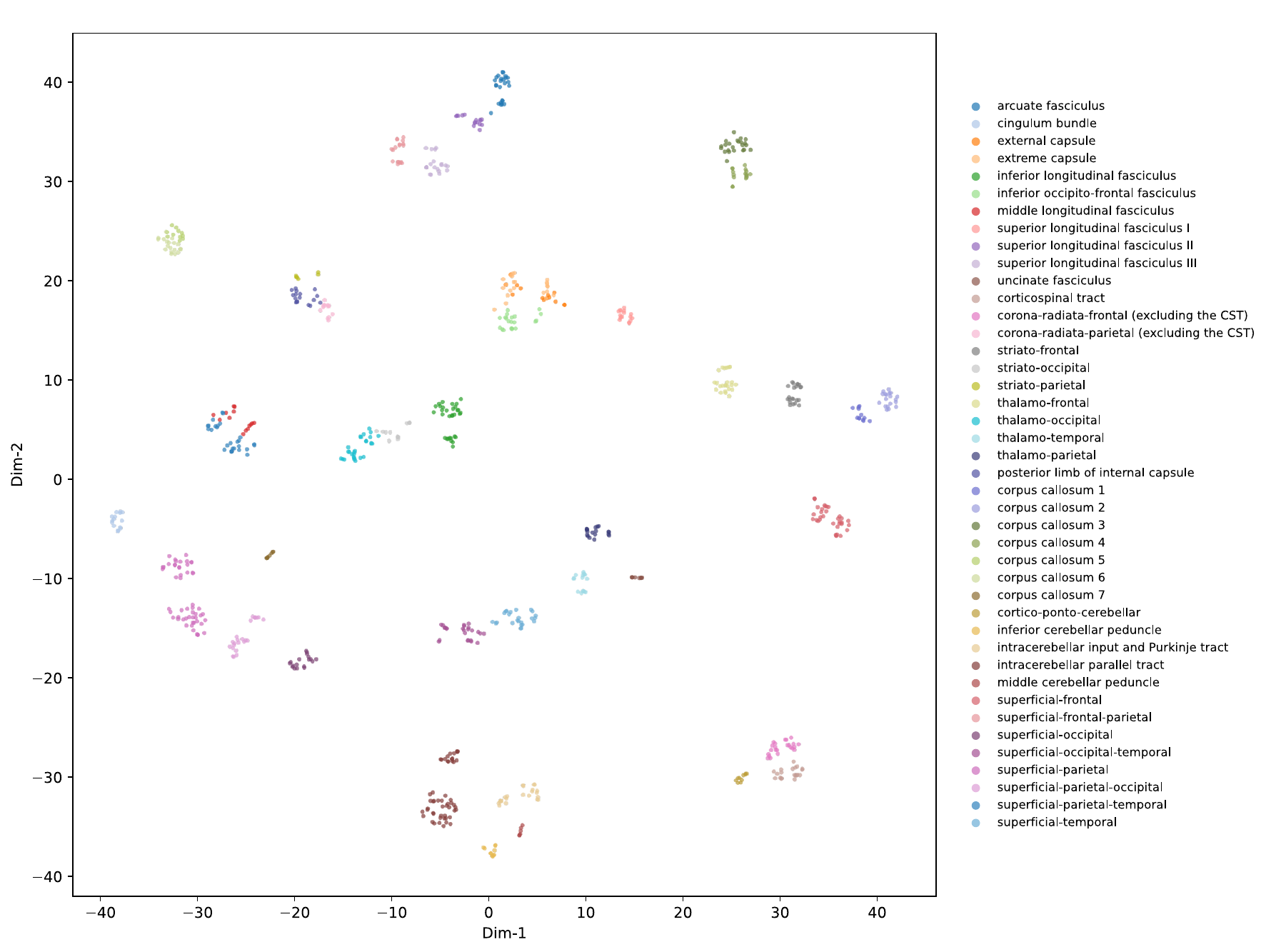}
    \caption{\textbf{t-SNE visualization of frozen contextualized streamline embeddings learned by TractFM.} 
    t-SNE projection of native 128-dimensional streamline embeddings extracted from an unseen test subject, comprising 842 streamlines from 42 representative tract classes. Each point denotes a single streamline and is colored by tract label. The embeddings show clear modular organization, with streamlines from the same tract forming compact neighborhoods and related pathways occupying nearby regions. The projected representation is consistent with strong tract separability and preservation of local neighborhood structure in the native embedding space.}
    \label{fig:tsne_embeddings}
\end{figure}

%% file: sec/misc/fig_attn.tex
\begin{figure}[t]
    \centering
    \includegraphics[width=\linewidth]{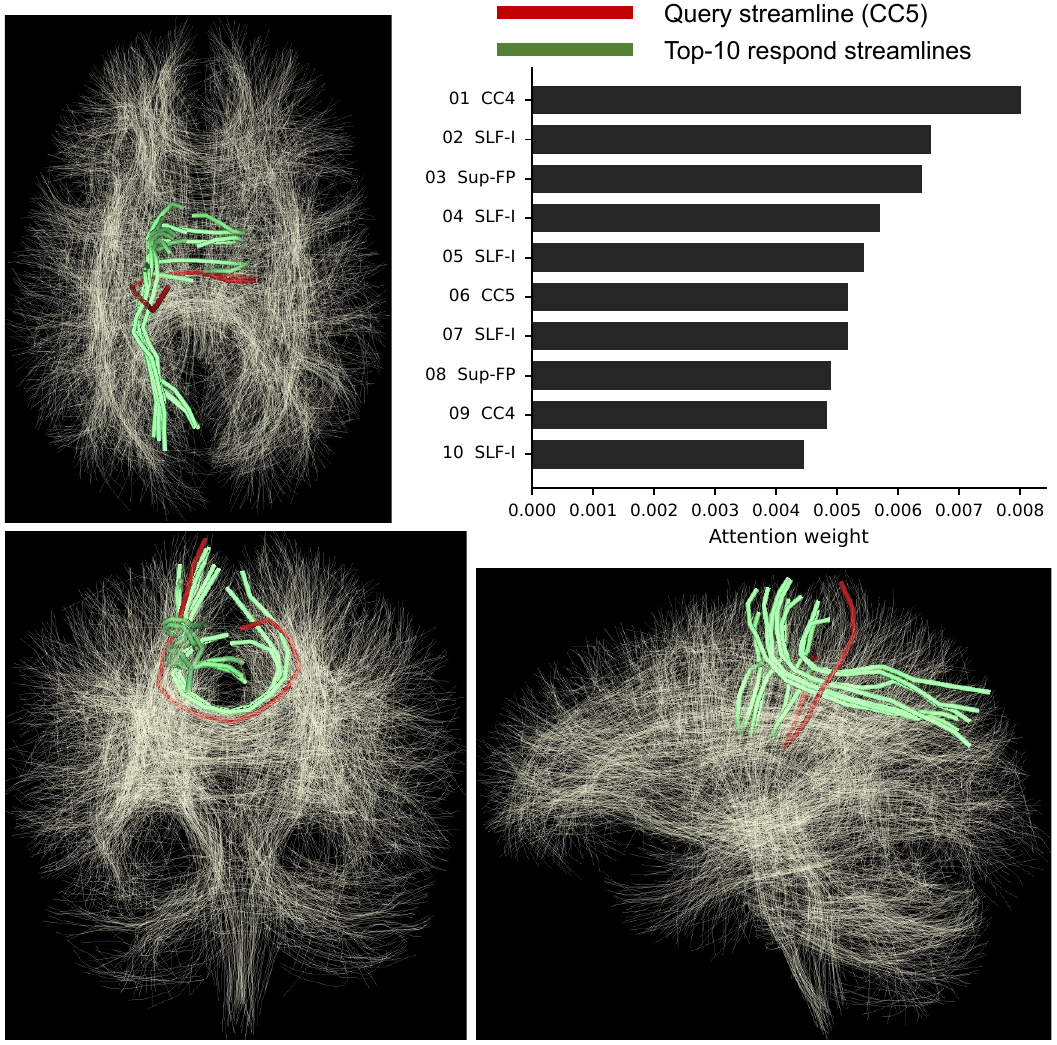}
    \caption{\textbf{Selective attention profile of TractFM for a representative corpus callosum streamline.} 
    Dense multi-head attention weights from the final tractogram encoder block are shown for a representative query streamline from corpus callosum 5 (CC5) in an unseen test example. Attention is sparse and highly non-uniform, concentrating on a restricted subset of contextual streamlines rather than being diffusely distributed across the tractogram. The strongest attended streamlines include both within-tract and cross-tract neighbors, notably from CC5, CC4, superior longitudinal fasciculus (SLF-I), and superficial frontal-parietal fibers (Sup-FP), indicating that TractFM integrates tract-specific local context with higher-order inter-tract information when encoding a target streamline.}
    \label{fig:attn_map}
\end{figure}

%% file: sec/misc/tab_downstream1.tex
\begin{sidewaystable}[p]
\centering
\caption{\textbf{Zero-tuning transfer performance of frozen TractFM embeddings across downstream cohorts.}
\textbf{a,} Best downstream performance achieved without task-specific fine-tuning using subject-level descriptors derived from frozen TractFM embeddings. Transfer was evaluated in the independent HCP-YA S1200 cohort and in the ABIDE-II+ADNI+CNP+PPMI cohort. For both cohorts, age prediction is reported by mean absolute error (MAE) and Pearson correlation coefficient ($r$), and sex classification by accuracy, area under the receiver operating characteristic curve (AUC), and F1 score.
\textbf{b,} Performance of different subject-level descriptors derived from frozen TractFM embeddings in the ABIDE-II+ADNI+CNP+PPMI cohort. For each descriptor, the best-performing downstream model for each task is shown. Lower MAE indicates better performance, whereas higher $r$, accuracy, AUC and F1 indicate better performance.}
\label{tab:zerotuning_transfer}

{\scriptsize
\renewcommand{\arraystretch}{1.08}

\begin{subtable}{\textheight}
\centering

\setlength{\tabcolsep}{5pt}
\begin{tabular}{@{}llllccccc@{}}
\toprule
Cohort & Task & Representation & Model & MAE $\downarrow$ & Pearson $r$ $\uparrow$ & Acc $\uparrow$ & AUC $\uparrow$ & F1 $\uparrow$ \\
\midrule
HCP-YA S1200 & Age & Mean+Std & Ridge   & \textbf{2.799} & \textbf{0.405} & --              & --              & --              \\
HCP-YA S1200 & Sex & Mean     & LogReg  & --             & --             & \textbf{0.889}  & \textbf{0.954}  & \textbf{0.889}  \\
ABIDE-II+ADNI+CNP+PPMI & Age & Mean+Std+CovEig+Entropy & SVR     & \textbf{5.360} & \textbf{0.962} & --              & --              & --              \\
ABIDE-II+ADNI+CNP+PPMI & Sex & Mean+Std+CovEig+Entropy & RBF-SVC & --             & --             & \textbf{0.763}  & \textbf{0.835}  & \textbf{0.813}  \\
\bottomrule
\end{tabular}
\caption{Best zero-tuning transfer performance across downstream cohorts}
\label{tab:zerotuning_overview}

\end{subtable}

\vspace{1.4em}

\begin{subtable}{\textheight}
\centering

\setlength{\tabcolsep}{4.5pt}
\begin{tabular}{@{}llccccccc@{}}
\toprule
\multirow{2}{*}{Representation} & \multirow{2}{*}{Dim.} & \multicolumn{3}{c}{Age prediction} & \multicolumn{4}{c}{Sex classification} \\
\cmidrule(lr){3-5}\cmidrule(lr){6-9}
& & Model & MAE $\downarrow$ & Pearson $r$ $\uparrow$ & Model & Acc $\uparrow$ & AUC $\uparrow$ & F1 $\uparrow$ \\
\midrule
Mean & 128 & SVR & 5.878 & 0.958 & RBF-SVC & 0.720 & 0.782 & 0.779 \\
Mean+Std & 256 & SVR & 5.440 & 0.962 & RBF-SVC & 0.760 & 0.826 & 0.812 \\
Mean+Max & 256 & SVR & 5.854 & 0.959 & RBF-SVC & 0.756 & 0.806 & 0.811 \\
CovEig & 128 & RF & 7.755 & 0.909 & RBF-SVC & 0.678 & 0.753 & 0.720 \\
Mean+CovEig & 256 & SVR & 5.650 & 0.959 & RBF-SVC & 0.744 & 0.807 & 0.798 \\
Mean+Std+CovEig+Entropy & 385 & SVR & \textbf{5.360} & \textbf{0.962} & RBF-SVC & \textbf{0.763} & \textbf{0.835} & \textbf{0.813} \\
\bottomrule
\end{tabular}
\caption{Performance of different subject-level descriptors in the ABIDE-II+ADNI+CNP+PPMI cohort}
\label{tab:multisite_feature_hierarchy}
\end{subtable}
}
\end{sidewaystable}

\begin{table*}[t]
\centering
\caption{Comparison of lightweight downstream models on the 385-dimensional Mean+Std+CovEig+Entropy descriptor in the ABIDE-II+ADNI+CNP+PPMI cohort.}
\label{tab:multisite_model_comparison}
\small
\begin{subtable}[t]{0.47\linewidth}
\centering
\begin{tabular}{lcc}
\toprule
Model & MAE $\downarrow$ & Pearson $r$ $\uparrow$ \\
\midrule
Ridge & 6.455 & 0.952 \\
SVR & \textbf{5.360} & \textbf{0.962} \\
RF & 5.927 & 0.955 \\
\bottomrule
\end{tabular}
\caption{Age prediction}
\label{tab:multisite_model_comparison_age}
\end{subtable}
\hfill
\begin{subtable}[t]{0.50\linewidth}
\centering
\begin{tabular}{lccc}
\toprule
Model & Acc $\uparrow$ & AUC $\uparrow$ & F1 $\uparrow$ \\
\midrule
LogReg & 0.726 & 0.817 & 0.773 \\
RBF-SVC & \textbf{0.763} & \textbf{0.835} & \textbf{0.813} \\
RF & 0.748 & 0.799 & 0.800 \\
\bottomrule
\end{tabular}
\caption{Sex classification}
\label{tab:multisite_model_comparison_sex}
\end{subtable}
\end{table*}

%% file: sec/misc/tab_robust.tex
\begin{table*}[t]
    \centering
    \caption{\textbf{Robustness of frozen TractFM downstream transfer to stochastic streamline subsampling and tractography method.}
    All experiments use the ABIDE-II+ADNI+CNP+PPMI cohort, the 385-dimensional Mean+Std+CovEig+Entropy descriptor, and frozen TractFM embeddings. Age prediction is evaluated with SVR; sex classification is evaluated with RBF-SVC. Lower MAE indicates better performance, whereas higher Pearson $r$, accuracy, AUC and F1 indicate better performance. The summary row reports mean (SD) across the three UKF samples.}
    \label{tab:robustness}
    
    \scriptsize
    \setlength{\tabcolsep}{3pt}
    \renewcommand{\arraystretch}{1.05}
    
    \begin{subtable}{0.7\linewidth}
    \centering
    \resizebox{\linewidth}{!}{
    \begin{tabular}{@{}lccccc@{}}
    \toprule
    \multirow{2}{*}{Run} & \multicolumn{2}{c}{Age prediction} & \multicolumn{3}{c}{Sex classification} \\
    \cmidrule(lr){2-3}\cmidrule(lr){4-6}
    & MAE $\downarrow$ & Pearson $r$ $\uparrow$ & Acc $\uparrow$ & AUC $\uparrow$ & F1 $\uparrow$ \\
    \midrule
    UKF sample 1 & 5.360 & 0.962 & 0.763 & 0.835 & 0.813 \\
    UKF sample 2 & 5.356 & 0.963 & 0.775 & 0.842 & 0.822 \\
    UKF sample 3 & 5.434 & 0.962 & 0.775 & 0.849 & 0.825 \\
    \midrule
    Mean (SD) & 5.383 (0.044) & 0.963 (0.001) & 0.771 (0.007) & 0.842 (0.007) & 0.820 (0.006) \\
    \bottomrule
    \end{tabular}
    }
    \caption{Repeated 10K-streamline samples using UKF tractography}
    \label{tab:robustness_sampling}
    \end{subtable}
    
    \vspace{1.2em}
    
    \begin{subtable}{\linewidth}
    \centering
    \begin{tabular}{@{}lccccc@{}}
    \toprule
    \multirow{2}{*}{Tractography method} & \multicolumn{2}{c}{Age prediction} & \multicolumn{3}{c}{Sex classification} \\
    \cmidrule(lr){2-3}\cmidrule(lr){4-6}
    & MAE $\downarrow$ & Pearson $r$ $\uparrow$ & Acc $\uparrow$ & AUC $\uparrow$ & F1 $\uparrow$ \\
    \midrule
    UKF & 5.360 & 0.962 & 0.763 & 0.835 & 0.813 \\
    DSI Studio & 5.715 & 0.958 & 0.737 & 0.820 & 0.778 \\
    iFOD2 & 5.050 & 0.967 & 0.764 & 0.844 & 0.817 \\
    \bottomrule
    \end{tabular}
    \caption{Alternative tractography methods}
    \label{tab:robustness_tractography}
    \end{subtable}
    \end{table*}

%% file: sec/discussion.tex
\section*{Discussion}
\label{sec:dis}

This study introduces TractFM, a foundation model that learns representations directly from whole-brain tractograms. By embedding all streamlines of a subject jointly, the pretrained model serves a dual purpose: it yields streamline-level embeddings for dense tract parcellation, and it aggregates into a reusable subject-level embedding. In multi-cohort, multi-protocol evaluations, whole-brain contextualization improved parcellation accuracy, and frozen subject-level embeddings transferred to chronological age and biological sex prediction without encoder fine-tuning. To our knowledge, TractFM is the first model to unify dense tract labeling and reusable subject-level representation learning within a single tractogram-level framework.

These results position TractFM in contrast to two directions in learning-based tractography analysis. One direction targets streamline parcellation, typically by classifying each streamline from local geometric descriptors or limited sampled contexts~\cite{xu2019objective,zhang2020deep,xue2023tractcloud,bisten2026rapidparc}; the other targets subject-level prediction, typically by feeding hand-designed tract-derived features into a task-specific classifier~\cite{mwangi2013prediction,he2022model,chen2023tractgraphcnn,chen2025tractgraphformer}. In both directions, the learned or engineered representations are tied to a single training objective and are not designed for reuse across tasks. Our findings are consistent with the broader move towards context-aware tractography modeling~\cite{xue2023tractcloud,bisten2026rapidparc,descoteaux2025millennium,zhang2025think}, and TractFM extends this trajectory by ingesting a full unordered tractogram in a single forward pass and by exposing a shared embedding space that supports both tract-level and subject-level use without retraining.

Beyond tractography, TractFM extends the foundation-model paradigm to a data structure that differs from the voxel-aligned 2D and 3D intensity grids that current medical-imaging foundation models primarily target~\cite{moor2023foundation,paschali2025foundation,pai2024foundation,wu2025towards,shao2025mri,zhang2025foundation}. The transferability of its embeddings suggests that this paradigm is applicable to imaging-derived geometric data more broadly and points towards analogous opportunities for other non-grid representations in connectomics and population neuroscience~\cite{descoteaux2025millennium,zhang2025think}.

TractFM is a starting point rather than an endpoint for tractogram foundation modeling, and several directions merit further study. The current model is pretrained with a single supervised objective; self-supervised strategies~\cite{khosla2020supervised,krishnan2022self,he2022masked} such as masked streamline modeling or cross-subject contrastive learning might enable pretraining on larger and more heterogeneous tractogram data. In addition, incorporating along-streamline microstructural measures might add a complementary information channel to the geometric inputs used here~\cite{zhang2022quantitative}. Finally, the downstream evaluations in this study cover two demographic phenotypes, and extension to clinical cohorts is a natural next step for assessing utility in disease characterization and outcome prediction~\cite{kao2019predicting,bonkhoff2022precision,gadot2024tractography}. Together, these directions position TractFM as a concrete foundation on which whole-brain tractogram representation learning can be developed.

%% file: sec/methods.tex
\section*{Methods}
\label{sec:method}

\subsection*{Problem formulation}
We conceptualize the diffusion MRI tractogram not as a collection of isolated streamlines, but as a unified, subject-level geometric entity. For a given subject, the input comprises an unordered set of $N$ streamlines. Each streamline is defined purely by its sequence of three-dimensional spatial coordinates. TractFM is designed to ingest this entire ensemble concurrently and produce dense predictions -- assigning a predefined anatomical label (from a space of $C$ classes) to every streamline in a single forward pass. 

Crucially, this whole-brain formulation ensures that the embedding of any individual streamline is contextualized by its geometric relationship to the rest of the brain's white-matter architecture. We use dense anatomical parcellation as the pre-training task because it provides supervision for every streamline and encourages the model to organize the tractogram according to whole-brain anatomy. We refer to TractFM as a foundation model because, after this pre-training stage, the encoder is reused as a frozen representation model across downstream datasets and phenotypic tasks. By contrast, prior tractography networks are typically optimized for a single labeling task rather than reused as general-purpose encoders.

\subsection*{Model architecture}
TractFM relies on a hierarchical architecture that progressively integrates local geometry and global brain context. The network comprises three core modules: a streamline encoder for local geometry, a tractogram encoder for whole-brain contextualization, and a task-specific projection head. A uniform embedding dimension $d$ is maintained throughout the network, yielding a subject-level contextualized embedding matrix $\bm{Z} \in \mathbb{R}^{N \times D}$.

\noindent\textbf{Input representation.}
For each subject, the input tractogram is uniformly sampled to a fixed set of $N$ streamlines, with each streamline resampled to $P$ equidistant points, yielding an input matrix $\bm{X} \in \mathbb{R}^{N \times 3 \times P}$.

\noindent\textbf{Local geometry extraction via the streamline encoder.}
The ordering of sampled points along a streamline is arbitrary: the same anatomical streamline may be represented with either endpoint first. To make the representation invariant to this reversal, we encode each streamline as an unordered set of 3D points using a PointNet-style symmetric encoder~\cite{qi2017pointnet}. For the $i$-th streamline, $\bm{X}_i \in \mathbb{R}^{3 \times P}$, shared $1\times1$ convolutions are applied independently to each point:
\begin{equation}
\bm{H}^{(1)}_i = \sigma\!\big(\mathrm{GN}\!(\bm{W}_1 * \bm{X}_i)\big),\quad
\bm{H}^{(2)}_i = \sigma\!\big(\mathrm{GN}\!(\bm{W}_2 * \bm{H}^{(1)}_i)\big),\quad
\bm{H}^{(3)}_i = \mathrm{GN}\!\big(\bm{W}_3 * \bm{H}^{(2)}_i\big),
\end{equation}
where $*$ denotes point-wise convolution, $\mathrm{GN}$ denotes Group Normalization, and $\sigma(\cdot)$ is the ReLU activation. We then apply symmetric max-pooling across the $P$ points to obtain a streamline-level embedding:
\begin{equation}
\bm{e}_i = \max_{p \in \{1,\ldots,P\}} \bm{H}^{(3)}_i(:,p) \in \mathbb{R}^{D}.
\end{equation}
Because the same transformation is applied to every point and the final pooling operation is permutation-invariant, reversing the order of points within a streamline does not change the resulting embedding. Since TractFM is trained on whole-subject tractograms with small effective batch sizes, we use Group Normalization in the streamline encoder. It normalizes groups of feature channels within each sample and is therefore well suited to tractogram-level training.

\noindent\textbf{Global contextualization via the tractogram encoder.}
The initial streamline encoder yields a matrix of local descriptors $\bm{E} =[\bm{e}_1;\ldots;\bm{e}_N] \in \mathbb{R}^{N \times d}$. To endow these representations with whole-brain anatomical context, we process $\bm{E}$ through a stack of Transformer-style self-attention blocks. Each block employs a pre-normalization residual architecture:
\begin{equation}
\tilde{\bm{E}} = \bm{E} + \mathrm{MHA}\!\big(\mathrm{LN}(\bm{E})\big), \qquad
\bm{E}' = \tilde{\bm{E}} + \mathrm{FFN}\!\big(\mathrm{LN}(\tilde{\bm{E}})\big),
\end{equation}
where $\mathrm{LN}$ denotes Layer Normalization, $\mathrm{MHA}$ is multi-head self-attention, and $\mathrm{FFN}$ is a position-wise feed-forward network with GELU activations and dropout. Notably, we omit positional encodings across the $N$ streamlines. This omission renders the tractogram encoder strictly permutation-equivariant, effectively modeling the tractogram as an unordered set and mathematically immunizing the network against arbitrary streamline sorting conventions introduced by tractography methods.

\noindent\textbf{Dense prediction and training objective.}
The embeddings contextualized by self-attention are projected to output classes via an MLP classifier head. Here, parcellation denotes dense anatomical labeling of every streamline in the tractogram into one of $C$ classes. This task provides supervision at streamline resolution while requiring the encoder to capture whole-brain anatomical context. For streamline $i$, the predicted logits are given by $\bm{\ell}_i = f_{\theta}(\bm{z}_i)$, where $\bm{z}_i \in \mathbb{R}^d$ is the output of the final self-attention block. The classifier comprises a sequence of Linear $\rightarrow$ LayerNorm $\rightarrow$ ReLU transformations, regularized by dropout ($p=0.3$) prior to the final projection to $C$ discrete classes. 

The training objective is the mean categorical cross-entropy over all $N$ streamlines within the tractogram:
\begin{equation}
\mathcal{L} = \frac{1}{N}\sum\nolimits_{i=1}^{N} \mathrm{CE}\!\left(\bm{\ell}_i, y_i\right),
\end{equation}
where $y_i \in \{1,\ldots,C\}$ denotes the atlas-derived ground-truth label, and the standard softmax transformation is computed implicitly within the loss function.

\noindent\textbf{Implementation details and computational trade-offs.}
In an inherently complex system like the human brain, a standard whole-brain tractogram typically comprises over $10^6$ streamlines. However, modeling interactions across millions of elements using standard global self-attention incurs an $O(N^2)$ memory and computational bottleneck. To make end-to-end whole-brain representation learning tractable on contemporary GPUs, we approximate the full tractogram by uniformly downsampling to $N=10{,}000$ streamlines during both training and inference. 

We empirically fixed the algorithmic hyperparameters to $P=15$ points per streamline, an embedding dimension of $d=128$, and $C=1600$ fine-grained parcellation classes. While computing over an approximated set of $10{,}000$ streamlines is a necessary pragmatic trade-off for current hardware limits, the formulation itself remains agnostic to $N$. As memory capacity and attention mechanisms improve, this leaves room to study denser tractograms, which may be important for analyses that require finer characterization of specific connections or networks.

\subsection*{Model training and evaluation protocols}

\noindent\textbf{Curated foundation dataset.}
For model development, we used a dataset of approximately one million labeled streamlines derived from a curated white-matter tractography atlas~\citep{zhang2018anatomically}. The atlas was constructed from 100 spatially registered whole-brain tractograms of young healthy adults from the Human Connectome Project (HCP)~\citep{van2013wu}. Streamlines were first grouped into fine-grained clusters using machine-learning methods, and the resulting clusters were then curated in collaboration with an expert neuroanatomist to provide high-confidence supervision. Labels are organized at two complementary granularities. At the coarse level, the atlas defines 43 tract categories: 42 anatomically meaningful whole-brain tracts and an additional ``other'' category that aggregates streamlines not consistent with the canonical anatomy, including anatomically implausible outliers. At the fine level, each streamline is assigned one of 1600 labels, and each fine class maps uniquely to one coarse tract category. This hierarchical labeling enables reporting of performance both at a clinically interpretable tract level (43 labels) and at a higher-resolution parcellation level (1600 labels).

\noindent\textbf{Data splits, model selection, and metrics.}
All splits are performed strictly at the subject level to prevent leakage, such that no streamline from a validation or test subject appears in training. Model selection is performed on a held-out validation set. Parcellation performance is reported using per-streamline top-1 accuracy and macro-averaged F1, with macro-F1 used as the primary selection criterion because it is sensitive to both frequent and rare classes. Final performance is reported on a held-out test set using the same evaluation protocol.

\noindent\textbf{Tractogram augmentation.}
Unless otherwise stated, we use geometric augmentation in both training and evaluation. For each subject, we generate 30 augmented views by sampling random three-dimensional transforms and applying each transform identically to all points in all streamlines of that subject. Each transform comprises a random rotation, translation, and mild anisotropic scaling, thereby perturbing global coordinate conventions while preserving the internal geometry and relative configuration of streamlines within the tractogram. During training, these 30 augmented views act as strong regularization against dataset-specific pose and scale. During evaluation, following~\citep{xue2023tractcloud}, we apply the same 30-view protocol and ensemble predictions across the 30 augmented views to improve robustness and reduce variance. 

\noindent\textbf{Optimization and training procedure.}
Models are implemented in PyTorch and trained using AdamW with weight decay. Because each forward pass processes an entire tractogram (all $N$ streamlines jointly), training is performed at the subject level. When GPU memory constrains the number of subjects processed per step, we use gradient accumulation to reach the desired effective batch size while preserving tractogram-level computation. Dropout is used in the self-attention blocks and in the classifier head for regularization. To enable tractogram-scale training, we use activation checkpointing within attention and feed-forward submodules to reduce memory consumption at the cost of additional recomputation.

\subsection*{Transfer learning for downstream phenotypic prediction}
\noindent\textbf{Rationale and downstream cohorts.}
To assess whether TractFM captures generalizable features of brain white-matter organization, we evaluated its ability to predict clinical phenotypes using only frozen spatial embeddings. We targeted chronological age (regression) and biological sex (binary classification). Downstream predictors were strictly precluded from fine-tuning the base TractFM weights; they were trained entirely on the subject-level descriptors derived from the pre-trained contextualized embeddings.

Validation was performed across two distinct domains: (i) the homogeneous HCP-YA S1200 cohort~\citep{van2013wu}, and (ii) a highly heterogeneous, multi-site pooled cohort comprising ABIDE-II~\citep{di2014autism}, ADNI~\citep{mueller2005alzheimer}, CNP~\citep{poldrack2016phenome}, and PPMI~\citep{marek2011parkinson}. The HCP cohort was divided into 7:1:2 train/validation/test partitions. For the multi-site evaluation, the 7:1:2 split was enforced independently within each constituent dataset prior to pooling, ensuring the preservation of multi-site variance across all partitions while maintaining absolute subject disjointness. 

\noindent\textbf{Deriving subject-level structural representations.} 
For a given subject, we passed the tractogram through the frozen network and extracted the output of the tractogram encoder, $\bm{Z} =[\bm{z}_1;\ldots;\bm{z}_N] \in \mathbb{R}^{N \times d}$. To reliably map this matrix of streamline embeddings into a fixed-dimensional global fingerprint, we defined a family of permutation-invariant projection functions, $\phi(\bm{Z})$. We comprehensively evaluated first-order statistics, boundary metrics, and second-order covariance summaries. Specifically, we computed the embedding mean ($\bm{m}$), embedding standard deviation ($\bm{s}$), and element-wise maximum ($\bm{a}$) across the $N$ streamline embeddings. To capture higher-order covariance structure of the embedding distribution, we computed the empirical covariance matrix $\bm{\Sigma} \in \mathbb{R}^{d\times d}$, together with its eigenvalue spectrum $\bm{\lambda}$ and spectral entropy $H(\bm{\lambda})$.

These operations permitted the systematic formulation of progressively richer subject representations. Our evaluation spanned a hierarchy of configurations: from the embedding mean alone, to first-order variability (mean + std) and boundary estimation (mean + max), up to richer statistical formulations that capture covariance structure (mean + std + eigenvalues + spectral entropy).

\noindent\textbf{Phenotype modeling and statistical baselines.} 
To rigorously bound the predictive utility of the TractFM embeddings, we evaluated them across a spectrum of lightweight supervised algorithms possessing different inductive biases~\cite{pedregosa2011scikit}. Age regression was modeled via ridge regression, support vector regression (SVR) with a radial basis function (RBF) kernel, and random forests. Sex classification utilized logistic regression, RBF-kernel support vector classifiers (SVC), and random forests. Final test performance additionally tracked Pearson's $r$ (with 95\% confidence intervals) and $R^2$ for regression, as well as the Area Under the Receiver Operating Characteristic Curve (AUC) and macroscopic F1-score for classification.